\documentstyle[12pt]{article}
\begin{document}
\setlength{\baselineskip}{.375in}
\vspace{0.7in}
\begin{center}
\Large\bf
Polylogarithm identities in a conformal field
theory\\
 in three dimensions\\
\vspace{0.75in}
\normalsize\rm
Subir Sachdev \\
\vspace{0.2in}
\em
Departments of Physics and Applied Physics, P.O. Box 2157,\\
Yale University, New Haven, CT 06520
\end{center}
\vspace{0.7in}
\rm
The $N=\infty$ vector $O(N)$ model is a solvable,
interacting field theory in three dimensions ($D$). In a recent paper
with A. Chubukov and J. Ye~\cite{self},
we have computed a universal number, $\tilde{c}$,
characterizing the size dependence of the
free energy at the conformally-invariant critical point of this theory.
The result~\cite{self} for $\tilde{c}$ can be
expressed in terms of polylogarithms.
Here, we use non-trivial polylogarithm
identities to show that $\tilde{c}/N = 4/5$, a rational number;
this result is curiously parallel to recent work on dilogarithm
identities in $D=2$ conformal theories.
The amplitude of the stress-stress correlator of this theory,
$c$ (which is the analog of the central charge),
is determined to be $c/N=3/4$, also rational.
Unitary conformal theories in $D=2$ always
have $c = \tilde{c}$; thus such a result is clearly not
valid in $D=3$.

\newpage
Consider a conformally-invariant field theory in $D$ dimensions.
Place it in a slab which is infinite in $D-1$ dimensions, but of
finite length $L$ in the remaining direction. Impose periodic boundary
conditions along this finite direction. An old result of Fisher and
de Gennes~\cite{degennes}
states that if hyperscaling is valid, the free energy
density ${\cal F} = -\log Z/V$ ($Z$ is the partition function and
$V$ is the total volume of the slab) satisfies
\begin{equation}
{\cal F} = {\cal F}_{\infty} - \frac{\Gamma [ D/2]
\zeta (D)}{ \pi^{D/2}} \frac{\tilde{c}}{L^D}.
\label{defpsi}
\end{equation}
Here ${\cal F}_{\infty}$ is the free energy density in the infinite system
and $\tilde{c}$ is a universal number. The coefficient of $1/L^D$ has
been chosen such that $\tilde{c} = 1$ for a single component, massless
free scalar field theory.
A similar parametrization has also been discussed
recently by Castro Neto and Fradkin~\cite{fradkin}.

A second universal number characterizing the conformal theory
is the amplitude of the two-point correlation function of the stress tensor
$T_{\mu\nu}$
in infinite flat space.
Cardy has proposed a normalization convention for $T_{\mu\nu}$
in arbitrary dimensions~\cite{cardyd3}, and shown that its two-point
correlation is of the following form
\begin{eqnarray}
\langle T_{\mu\nu} (r) T_{\lambda \sigma} (0) \rangle &=& \frac{c}{r^{2D}}
\left[ \left( \delta_{\mu\lambda} - \frac{2r_{\mu} r_{\lambda}}{r^2}
\right) \left( \delta_{\nu\sigma} - \frac{2r_{\nu} r_{\sigma}}{r^2}
\right) +
\left( \delta_{\mu\sigma} - \frac{2r_{\mu} r_{\sigma}}{r^2}
\right) \left( \delta_{\nu\lambda} - \frac{2r_{\nu} r_{\lambda}}{r^2}
\right) \right.\nonumber\\
&~&~~~~~~~~~~~~~~~~~~~~~~~~~~~~~~~~
\left.- \frac{2}{D} \delta_{\mu\nu} \delta_{\lambda\sigma} \right].
\label{stress}
\end{eqnarray}
This defines a universal amplitude, $c$, which is the analog of the
central charge in $D=2$ conformal field theories.
A key property of $D=2$ unitary conformal field theories is
$\tilde{c}=c$~\cite{affleck}. The generalization of this result
to arbitrary $D$, and in particular $D=3$, remains an important open
problem.

It would clearly be interesting to obtain results for $\tilde{c}$ and $c$
for specific models
in dimensions other than $D=2$.
In a recent paper by A. Chubukov, myself and J. Ye~\cite{self}
on the critical properties of
two-dimensional quantum antiferromagnets,
the value of $\tilde{c}$ was computed for the vector $O(N)$ model in
$D=3$ in a $1/N$ expansion. In this note we highlight some features of the
computation of $\tilde{c}$ at $N=\infty$, as we believe the results
may be of interest to a broader audience of conformal field theorists.
The result for $\tilde{c}$ at $N=\infty$ can be
expressed in terms of di- and trilogarithm functions. Below, we use
some known polylogarithm identities to simplify the result for
$\tilde{c}$. The appearance of these identities is surprisingly
parallel to recent work~\cite{nahm}
establishing a connection between dilogarithmic
identities and the rational
central charge of $D=2$ conformal theories.
We will also compute the value of $c$ at $N=\infty$ in the $D=3$
$O(N)$ model.

We consider the field theory with the action
\begin{equation}
S = \frac{N}{2g} \int d^3 x ( \partial n )^2
\label{action}
\end{equation}
where $n$ is a $N$-component real vector of unit length, $n^2 = 1$.
The fixed length constraint is actually not crucial and identical
universal properties can be obtained in a soft-spin theory with a $n^4$
interaction term~\cite{edouard}.
The theory has to be suitably regulated in the ultraviolet
by a momentum cutoff $\Lambda$. It becomes conformally invariant
at a critical value $g = g_c = \alpha / \Lambda$ which separates the
$g < g_c$ Goldstone  phase with broken $O(N)$ invariance, from
the $g > g_c$ massive phase. The location of the critical point, $\alpha$,
is of course non-universal and will depend upon the cutoff scheme.

The formal structure of the $N\rightarrow \infty$ limit is quite standard.
The fixed length constraint is imposed by an auxilliary field $\lambda$.
After integrating out the $n$ field, the $N=\infty$ theory is given by the
saddle point of the resulting functional integral. In this manner we find
\begin{equation}
\frac{{\cal F}}{N} = \frac{1}{2}
\mbox{Tr} \log ( - \partial^2 + m^2 ) - \frac{m^2}{2 g}
\label{fovern}
\end{equation}
where $m^2$ is the saddle-point value of $\lambda$. The critical point is at
$g=g_c$, where
\begin{equation}
\frac{1}{g_c} = \int \frac{d^3 p}{8 \pi^3} \frac{1}{p^2},
\end{equation}
and $m^2 = 0$ in the infinite volume system. In the slab with thickness
$L$, however, we find at $g=g_c$ that
\begin{equation}
m = m_L = \frac{ 2 \log \tau}{L},
\end{equation}
where $\tau = ( \sqrt{5} + 1 )/2$ is the golden mean.

To compute $\tilde{c}$,
we now need to evaluate the $\mbox{Tr} \log$ in (\ref{fovern}) in the
slab geometry.
The momentum along the finite direction is quantized in integer multiples
of $2 \pi / L$. The summation over these discrete modes can be accomplished
with the identity
\begin{equation}
\lim_{M \rightarrow \infty} \left[ \frac{1}{L} \sum_{n=-M}^{M}
\log \left( \frac{4 \pi^2 n^2}{L^2} + a^2 \right)
- \int_{-2\pi M/L}^{2\pi M/L} \frac{d \omega}{2 \pi} \log ( \omega^2
+ a^2 )\right] = \frac{2}{L} \log\left(1- e^{-L|a|} \right),
\end{equation}
where $a$ is any constant.
The expression for ${\cal F}$ in the slab of width $L$ at $g=g_c$ is then
easily shown to be
\begin{equation}
\frac{{\cal F}}{N} = \frac{1}{L} \int \frac{d^2 k}{4 \pi^2} \log \left( 1 -
e^{-L\sqrt{m_L^2 + k^2}} \right)
+ \frac{1}{2} \int \frac{d^3 p}{8 \pi^3} \left[
\log(p^2 + m_L^2) - \frac{m_L^2}{p^2} \right]
\end{equation}
The second integral is of course badly divergent in the ultraviolet.
All divergences however disappear after the infinite volume result
has been subtracted, in which case
\begin{equation}
\frac{{\cal F}- {\cal F}_{\infty}}{N} =
\frac{1}{L} \int \frac{d^2 k}{4 \pi^2} \log \left( 1 -
e^{-L\sqrt{m_L^2 + k^2}} \right)
+ \frac{1}{2} \int \frac{d^3 p}{8 \pi^3} \left[
\log\left(\frac{p^2 + m_L^2}{p^2}\right) - \frac{m_L^2}{p^2} \right]
\label{ff}
\end{equation}
These integrals can be expressed in terms of polylogarithms.
We will skip the straightforward intermediate steps and
present our final result
for $\tilde{c}$ obtained from (\ref{ff}) and (\ref{defpsi})
\begin{equation}
(1/N){\rm Li}_3 (1) \tilde{c} = {\rm Li}_3 (2 - \tau)
- \log( 2 - \tau ) {\rm Li}_2 ( 2 - \tau ) - \frac{1}{6} \log^3 ( 2 - \tau )
\label{psi}
\end{equation}
where $2 - \tau = 1/\tau^2 = (3 - \sqrt{5})/2$, and the polylogarithm
function is defined by analytic continuation of the series
\begin{equation}
{\rm Li}_p ( z) = \sum_{n=1}^{\infty} \frac{z^n}{n^p}.
\end{equation}
Note ${\rm Li}_p ( 1) = \zeta (p)$.

Remarkably, it turns out that $2-\tau$ is one of only
three real, positive, $z$
for which both ${\rm Li}_2 (z)$ and ${\rm Li}_3 (z)$ can be expressed in
terms of elementary functions~\cite{lewin}
(the other points are $z=1$ and $z=1/2$).
As shown in the book by Lewin~\cite{lewin},
the value
of ${\rm Li}_2 ( 2 - \tau )$ follows from a combined analysis of the following
identities
\begin{eqnarray}
{\rm Li}_2 (z) + {\rm Li}_2 ( 1 - z) &=& \frac{\pi^2}{6} - \log z \log(1-z)
\nonumber \\
{\rm Li}_2 (z) + {\rm Li}_2 \left(\frac{-z}{1 - z}\right)
&=&  - \frac{1}{2}\log^2 (1-z)
\nonumber \\
\frac{1}{2}{\rm Li}_2 (z^2) + {\rm Li}_2 \left(\frac{-z}{1 - z}\right)
-{\rm Li}_2 (-z)&=&  - \frac{1}{2}\log^2 (1-z)
\end{eqnarray}
To see the special role of the golden mean in these identities,
note that two of the arguments $z^2$ and $-z/(1-z)$ coincide
when $z^2 - z - 1=0$. The solutions of this are $z=\tau, 1-\tau$.
It is not difficult to show that
the above identities evaluated at $z=2-\tau$, $\tau-1$, and $1-\tau$, can be
combined to uniquely determine
${\rm Li}_2 ( 2 -\tau)$~\cite{lewin}:
\begin{equation}
{\rm Li}_2 (2-\tau) = \frac{\pi^2}{15} - \frac{1}{4} \log^2 (2-\tau )
\label{li2}
\end{equation}
Similarly, the identities~\cite{lewin}
\begin{eqnarray}
\frac{1}{4}{\rm Li}_3 (z^2 ) &=&  {\rm Li}_3 (z) + {\rm Li}_3 (-z)
\nonumber \\
{\rm Li}_3 (z) + {\rm Li}_3 \left(\frac{-z}{1 - z}\right)
+{\rm Li}_3 (1-z)&=& {\rm Li}_3 (1)+ \frac{\pi^2}{6} \log(1-z)
\nonumber \\
&~&~~~~~- \frac{1}{2}\log z \log^2 (1-z) + \frac{1}{6} \log^3 ( 1-z)
\end{eqnarray}
evaluated at $z=2-\tau$ and $\tau-1$
yield~\cite{lewin}
\begin{equation}
{\rm Li}_3 (2-\tau) = \frac{4}{5} {\rm Li}_3 ( 1)
+ \frac{\pi^2}{15}\log(2-\tau) - \frac{1}{12} \log^3 (2-\tau )
\label{li3}
\end{equation}
Inserting (\ref{li2}) and (\ref{li3}) into (\ref{psi}),
we get one of our main results
\begin{equation}
\frac{\tilde{c}}{N} = \frac{4}{5}
\end{equation}
Surprisingly, $\tilde{c}/N$ has turned out to be a rational number,
although none of the intermediate steps suggested that this might be
the case. Interestingly, this phenomenon is similar to
that in recent determinations of
$\tilde{c}$ from the size dependence of ${\cal F}$
in $D=2$ conformal theories~\cite{affleck,nahm}. There, the free energy
was determined from
integrable lattice models, or by evaluating the characters of a
representation
of the Virasoro algebra; in both cases the result was obtained in
terms of dilogarithm sums, which thus must equal the rational central
charge.

We turn next to the determination of $c$ for the $D=3$, $N=\infty$
vector $O(N)$ model. The stress tensor $T_{\mu\nu}$ for
(\ref{action}) is
\begin{equation}
T_{\mu\nu} = \frac{4\pi}{g} \left(
\partial_{\mu} n \partial_{\nu} n - \frac{\delta_{\mu\nu}}{2}
( \partial n)^2 \right) - \delta_{\mu\nu} t
\end{equation}
where $t$ is a cutoff-dependent subtraction needed to
make $T_{\mu\nu}$
a proper scaling variable at the critical point.
The general structure of these subtractions in the $1/N$ expansion
for arbitrary operators in the $O(N)$ model with a hard momentum cutoff
has been discussed
by Ma~\cite{ma}. Here, we simply note that dimensional regularization of
the loop integrals in the vicinity of $D=3$ leads to $t = 0$
at $N=\infty$. We evaluated $\langle T_{\mu\nu} ( r )
T_{\lambda\sigma} (0) \rangle$ at $N=\infty$
in the infinite system
using dimensional regularization. There are two Feynman graphs which
contribute at this order~\cite{ma}, including one involving
fluctuation of the auxilliary field, $\lambda$, which
imposed the fixed length
constraint. The loop integrals are quite tedious, but straightforward.
We found that our final result was indeed
consistent with (\ref{stress}) with
\begin{equation}
\frac{c}{N} = \frac{3}{4}
\end{equation}
Note that $c \neq \tilde{c}$, unlike $D=2$. Instead, we have
$c/\tilde{c} = 15/16$ in this theory.

We emphasize that all of the results of this paper are special to
$D=3$; $\tilde{c}$ can also be computed for general $D$, but the
results simplify only in $D=3$.
The major question raised by this work is, of course, whether
$\tilde{c}$ and $c$ have any of these
special properties at finite $N$ in $D=3$.
It would also be interesting to obtain the simple $D=3$ results for $c$ and
$\tilde{c}$ at $N=\infty$ by algebraic methods.

I thank A. Chubukov, G. Moore, N. Read and R. Shankar for useful discussions.
This research
was supported
by NSF Grant No. DMR 8857228.


\begin{thebibliography}{99}
\bibitem{self} A.V. Chubukov, S. Sachdev and J. Ye, paper 9304046 on
cond-mat@babbage.sissa.it.
\bibitem{degennes} M.E. Fisher and
P.-G. de Gennes, C.R. Acad. Sci. Ser. B {\bf 287},
207 (1978); V. Privman and M.E. Fisher, Phys. Rev. B {\bf 30},
322 (1984).
\bibitem{fradkin} A.H. Castro Neto and E. Fradkin, paper 9301009 on
cond-mat@babbage.sissa.it
\bibitem{cardyd3} J.L. Cardy, Nucl. Phys. {\bf B290}, 355 (1987).
\bibitem{affleck} H.W.J. Blote, J.L. Cardy and M.P. Nightingale,
Phys. Rev. Lett. {\bf 56}, 742 (1986); I. Affleck, Phys. Rev. Lett.
{\bf 56}, 746 (1986).
\bibitem{nahm} V.V. Bazhanov and N. Yu. Reshetikhin,
Int. J. Mod. Phys. A {\bf 4}, 115 (1989); A.N. Kirillov, J. Sov. Math.
{\bf 47} 2450 (1989); T.R. Klassen and E. Melzer, Nucl. Phys. B
{\bf 338}, 485 (1990) and {\bf 370}, 511 (1992);
A. Klumper and P.A. Pearce, J. Stat. Phys. {\bf 64}, 13 (1991);
F. Ravanini, Phys. Lett. B {\bf 282}, 73 (1992);
W. Nahm, A. Recknagel, and M. Terhoeven, paper 9211034 on
hep-th@xxx.lanl.gov.
\bibitem{edouard} E. Brezin and J. Zinn-Justin,
Phys. Rev. B {\bf 14}, 3110 (1976).
\bibitem{lewin} {\em Polylogarithms and Associated Functions}, by
L. Lewin, North Holland, New York (1981).
\bibitem{ma} S.-k. Ma, Phys. Rev. {\bf 10}, 1818 (1974).
\end{thebibliography}
\end{document}